\documentclass[a4paper,english,aps,prx, twocolumn, floatfix, showpacs, superscriptaddress]{revtex4-1}

\usepackage{amsmath}
\usepackage{amssymb}
\usepackage{graphicx}
\usepackage{hyperref}
\usepackage{color}

\newcommand{\figref}[1]{Fig.~\ref{#1}}

\begin{document}

\title{Statistical Analysis of the Chern Number in the interacting Haldane-Hubbard Model}
\author{Thomas Mertz}
\author{Karim Zantout}
\author{Roser Valent\'{i}}
\affiliation{Institut f\"ur Theoretische Physik, Goethe-Universit\"at, 60438 Frankfurt am Main, Germany}

\begin{abstract}
In the context of many-body interacting systems described by a topological Hamiltonian, we investigate the robustness of the Chern number with respect to different sources of error in the self-energy.
In particular, we analyze the importance of non-local (momentum dependent) vs. local contributions to the self-energy and show that the local self-energy provides a qualitative description of the topological phase diagrams of many-body interacting systems, whereas the explicit momentum-dependence constitutes a correction to the exact location of the phase transition. For the latter, we propose a statistical analysis, on the basis of which we develop a stochastic upper bound for the uncertainty of the Chern number as a function of the amount of momentum-dependence of the self-energy. We apply this analysis to the Haldane-Hubbard model and discuss the implications of our results for a general class of many-body interacting systems.
\end{abstract}

\maketitle

\section{Introduction}
Since the discovery of the integer quantum Hall effect
\cite{vonKlitzing1986,Laughlin1981} topology has been considered a key
ingredient in characterizing phases of matter, in particular through the
formulation of topological order parameters. 
The topology of a system can be
characterized by both bulk and surface properties of a sample. The latter is
reflected in the topological surface states \cite{Halperin1982,Rammal1983} 
at the interfaces between topologically inequivalent crystals by
virtue of the bulk-boundary correspondence \cite{Hatsugai1993}. On the other hand, 
the bulk
properties are typically characterized in terms of topological invariants
\cite{Thouless1982} which define an equivalence relation among the set of
non-interacting Hamiltonians.

In integer quantum Hall systems the topological invariant is given by the Chern number \cite{Thouless1982}
\begin{equation}
\begin{split}
	C = \frac{i}{2\pi}\sum_n\iint \mathrm{d}^2k \Big( &\partial_{k_y}\langle k, n|\partial_{k_x} | k, n\rangle \\
	- &\partial_{k_x}\langle k, n|\partial_{k_y} | k, n\rangle\Big) ,
\end{split}
	\label{eq:chern_invariant}
\end{equation}
defined as the integral over the Berry curvature.
Since  the Berry phase is the phase acquired by an electron on a path around the Brillouin zone,
 it is clear that the Chern number primarily describes the momentum-dependence of the Hamiltonian.
The definition given in Eq.~\eqref{eq:chern_invariant} applies, in principle, only to single-electron systems, where the Bloch theorem guarantees the existence of eigenstates $|k, n\rangle$, with quasi-momentum $k$ and band index $n$.

In recent years, a lot of effort has been devoted to understand the topological
properties of non-interacting systems \cite{Fu2007,Fu2011,Wan2011,Bradlyn2017,Rachel2018} and most recent advances include, for
instance, the prediction of higher order topological insulators protected by
spatial symmetries \cite{Schindler2018,Trifunovic2019}. The progress for
interacting systems has been more difficult due to the challenges posed by the
many-body nature of the interactions. Nonetheless, a few important results have
been obtained in the past. 

\begin{figure}
        \includegraphics[scale=1]{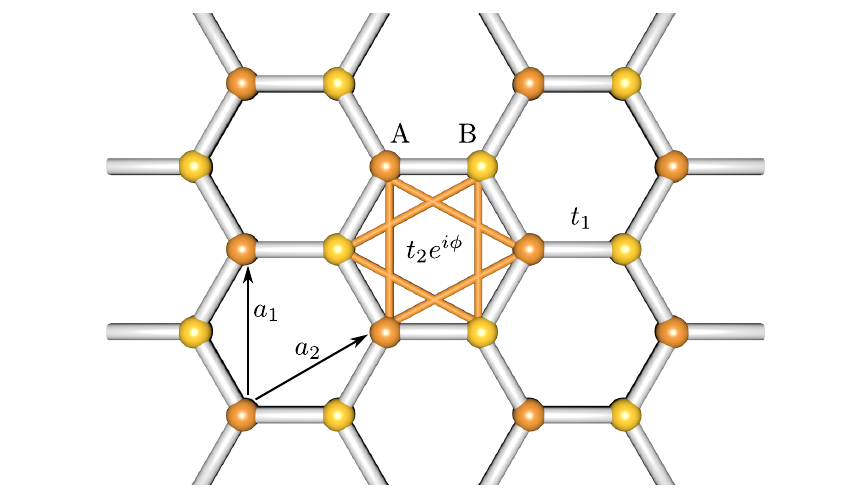}
        \caption{Illustration of the Haldane model, which includes nearest-neighbor hopping $t_1$, next-nearest neighbor hopping $t_2$ with phase $\pm\phi$ for (anti-) clockwise hopping. The two sublattices of the honeycomb lattice are offset by a mass $\Delta$. $a_1,a_2$ are the lattice vectors.}
        \label{fig:haldane_illustration}
\end{figure}

It has been shown that the Hall conductivity of an interacting system \cite{Ishikawa1986} can be computed through Eq.~\eqref{eq:chern_invariant} by using a many-body formalism based on Green's functions \cite{Wang2012,Wang2013}.
In this approach one replaces the Hamiltonian with a convenient effective topological Hamiltonian, which is defined by
\begin{equation}
	H_t(k) = H_0(k) + \Sigma(k, \omega=0),
	\label{eq:topological_hamiltonian_def}
\end{equation}
where $H_0(k)$ is the non-interacting single-particle Hamiltonian and $\Sigma(k,\omega)$ is the self-energy of the original interacting Hamiltonian.
Eq.~\eqref{eq:topological_hamiltonian_def} can be reinterpreted as an effective model, where one adds an additional potential---in this case the self-energy---such that it describes the correct Chern invariant for the interacting system. 
This approach is valid if a smooth connection to the zero-frequency limit can be established as is the case away from the Mott insulating phase.

In the past, many studies of topological models have neglected the momentum dependence of the self-energy by using the popular dynamical mean-field theory \cite{Vanhala2016,Kumar2016,Budich2013} and only very few results are available using approximate non-local methods, e.g.~\cite{Budich2012,Tupitsyn2019}. This seems paradoxical, as the dispersion of the self-energy is expected to be a key ingredient in the computation of the Chern number [Eq.~\eqref{eq:chern_invariant}].

Motivated by this paradox and the fact
that there is no guarantee that self-energies available through approximate methods produce the correct  Chern number, we investigate here how local and non-local contributions to the self-energy are responsible for the determination of the topological invariant. For that we propose a method based on a statistical analysis of the self-energy, that (i) does not require an a priori knowledge of the correct self-energy and (ii) explores a large phase space of possible self-energies and therefore is general enough to allow for \emph{universal} statements on the nature of topological phases of interacting systems. 
In the following we introduce the method and consider the Haldane-Hubbard model as a testbed for assessing its validity and predictive power.
Our analysis shows that, albeit the intrinsic momentum-dependent definition of the Chern number,
non-local contributions to the self-energy add only a small uncertainty to the effects of the local self-energy in
interacting systems described by topological Hamiltonians.

\subsection{Haldane-Hubbard Model}

We study the Haldane-Hubbard model at half-filling on the honeycomb lattice, cf.~\figref{fig:haldane_illustration}, which combines Haldane's model for the integer quantum Hall effect \cite{Haldane1988} with a local Hubbard interaction of the form
\begin{equation}
	H = \sum_k (c_{A}^\dagger, c_{B}^\dagger) h_k (c_{A}, c_{B})^T + U \sum_i n_{i\uparrow} n_{i\downarrow},
	 \label{eq:haldane_hamiltonian}
\end{equation}
with
\begin{equation}
\begin{split}
	h_k = &2t_2 \cos\phi \left[ \cos(k_1) + \cos(k_2) + \cos(k_2-k_1)\right] \sigma_0 \\
	&+ t_1\left[ 1 + \cos(k_2) + \cos(k_2-k_1)\right] \sigma_1 \\
	&- t_1 \left[\sin(k_2) + \sin(k_2-k_1)\right] \sigma_2 \\
	&+ \big[ \Delta - 2t_2 \sin\phi [ \sin(k_1) + \sin(k_2) \\
	&\quad+ \sin(k_2-k_1)] \big]\sigma_3,
\end{split}
\label{eq:haldane_hamiltonian_matrix}
\end{equation}
where $A/B$ stand for sublattice indices (see \figref{fig:haldane_illustration}), $t_1,t_2$ are the nearest and
 next-nearest neighbor hopping amplitudes, respectively, $\phi$ is the phase associated with the 
next-nearest neighbor hopping, $\Delta$ a trivial mass term and $\sigma_i$ are the Pauli matrices in sublattice space.
Throughout this article, we keep $t\equiv t_1$, $t_2/t_1=0.2$ and $\phi=\pi/2$ fixed. For this set of parameters, the Haldane model ($U=0$) has a topological phase transition from a topological insulator to a trivial band insulator at $\Delta_c\approx 1.04 t$.

\begin{figure}
	\includegraphics[scale=1]{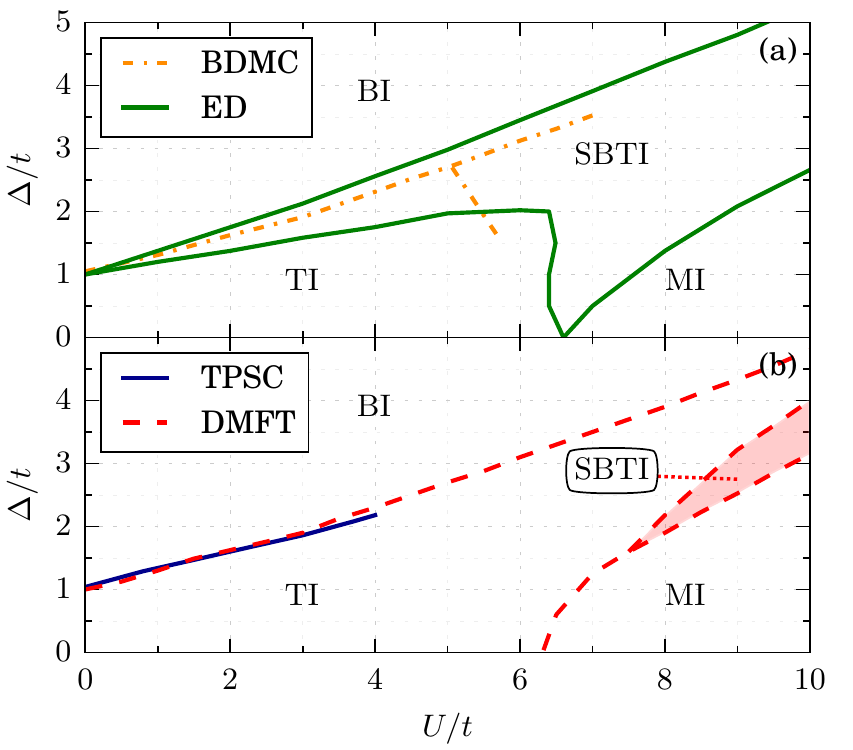}
	\caption{Phase diagram of the Haldane Hubbard model. In addition to our TPSC calculations we show for comparison the ED and DMFT data from \cite{Vanhala2016} and BDMC from \cite{Tupitsyn2019}. In (a), BDMC (orange line), ED (green line), and in (b), TPSC (blue line), DMFT (red line) are shown. In DMFT the $C=1$ (SBTI) phase extends only down to a finite minimal value of $\Delta$ (red colored area), while it survives down to $\Delta =0$ for ED and BDMC. Qualitatively, the ED and BDMC phase diagrams are similar, except for ED predicting an SBTI phase at $U\rightarrow 0$.}
	\label{fig:phase_diagram}
\end{figure}

The phase diagram of the Haldane-Hubbard model at half filling has been studied extensively in recent years \cite{Vanhala2016,Imriska2016,Tupitsyn2019} by a variety of methods including static mean-field theory (MF), dynamical mean-field theory (DMFT), exact diagonalization (ED), dynamical cluster approximation and quantum Monte Carlo approaches.
In \figref{fig:phase_diagram} we recapitulate the current understanding of the phase diagram from the contemporary literature and include our results obtained with the Two-Particle Self-Consistent (TPSC) technique \cite{Vilk1994,Zantout2018} for low to intermediate values of the on-site interaction $U$ where the method is most reliable (\figref{fig:phase_diagram}(b), blue line). The phases observed are a topological insulator (TI) with $C=2$ (both spins have Chern number 1) at low $\Delta$ and $U$, a trivial band insulator (BI) with $C=0$ at large $\Delta$, a Mott insulator (MI) at large $U$ and an SU$(2)$ symmetry-broken topological insulator (SBTI) with $C=1$ at intermediate values.
The TPSC calculations are in good agreement with DMFT \cite{Vanhala2016} and
Bold Diagrammatic Monte Carlo (BDMC) \cite{Tupitsyn2019} in the regions of $U$
studied.

In DMFT (\figref{fig:phase_diagram}(b), red line) the location of the TI$\rightarrow$SBTI phase transition
strongly depends on the value of $\Delta$, and approaches the TI$\rightarrow$BI
transition line asymptotically, while recent BDMC calculations suggest the
existence of a critical point where the two lines intersect (\figref{fig:phase_diagram}(a), dotted blue line). In order to rule
out the possibility of this discrepancy being a consequence of  different
simulation protocols, we have performed DMFT calculations using the protocol
laid out in Ref.~\cite{Tupitsyn2019} and confirmed the previously published
DMFT data \cite{Vanhala2016}. The shift of the TI$\rightarrow$SBTI transition
to lower values of $U$ in BDMC with respect to DMFT means that the interacting system
obtains magnetic order sooner than DMFT predicts, which seems to be an
indicator of strong non-local contributions to the self-energy.

\section{Chern Number Analysis: local contributions}
In order to settle the origin of agreement and discrepancies among the various approaches and to establish which contributions in
 the self-energy influence the nature of the topological phases, we introduce in what follows
a detailed analysis of the calculation of the Chern number through the self-energy as defined
in Eq.~\eqref{eq:topological_hamiltonian_def}.

As shown in Ref.~\cite{Mertz2018}, one can decompose the self-energy into a local part $\Sigma_\mathrm{loc}$ and a non-local part $\Sigma_\mathrm{non-loc}$
  as
\begin{equation}
	\Sigma(k, \omega) = \Sigma_\mathrm{loc}(\omega) + \Sigma_\mathrm{non-loc}(k, \omega),
	\label{eq:sigma_loc_nonloc}
\end{equation}
where $\Sigma_\mathrm{non-loc}(k, \omega)$ has a vanishing momentum average, i.e.~corrections to the local self-energy are already absorbed in $\Sigma_\mathrm{loc}$.
In order to quantify the explicit momentum-dependence in $\Sigma_\mathrm{non-loc}(k, \omega)$
 we define the self-energy dispersion amplitude \cite{Mertz2018}
\begin{equation}
	d_a(\omega) = \max_{k,k^\prime} \parallel \Sigma(k, \omega) - \Sigma(k^\prime, \omega)\parallel.
\label{eq:da}
\end{equation}
Since only the zero-frequency self-energy enters the topological Hamiltonian [Eq.~\eqref{eq:topological_hamiltonian_def}] we only have to consider the physics associated with $d_a(\omega=0)$. Therefore, hereafter, we use the shorthand notation $d_a \equiv d_a(0)$.

\subsection{SU$(2)$-symmetric self-energy}
\label{sec:local}

We first focus on the local self-energy. More specifically
we will analyze the effects of the diagonal and off-diagonal components
of the local self-energy on the Chern number.

In the Hartree approximation (mean field (MF)) the local self-energy is given by
\begin{equation}
	\Sigma^\mathrm{MF} = \frac{U}{2} n.
\end{equation}
For Hamiltonians in a bipartite lattice
with a mass term $\Delta$ the density alternates between $A/B$ sublattices (see \figref{fig:haldane_illustration}), such that upon addition of a constant term the self-energy can be written as
\begin{equation}
	\Sigma^\mathrm{MF} = -\frac{U}{2} \delta n \;\sigma_3 + \mathrm{const.},
\end{equation}
where $\delta n = (n_B-n_A)/2$ and $\sigma_3$ is the third Pauli matrix. 
The constant is absorbed in the chemical potential. Therefore, within the Hartree approximation, $\Sigma$ \emph{reduces} the strength of the mass term with respect to the non-interacting contribution (Eq.~\eqref{eq:topological_hamiltonian_def}) and the topological transition shifts to larger $\Delta$ with increasing $U$.

We can easily see that the above Hartree argument is exact for the local contribution of the self-energy. The self-energy generally obeys the symmetry
\begin{equation}
	\Sigma_A(k, \omega) = -\Sigma_B(k, \omega),
	\label{eq:sigma_diagonal_symmetry}
\end{equation}
which follows from the symmetry of the Hamiltonian, Eq.~\eqref{eq:haldane_hamiltonian},
up to a constant term, which we can neglect as it is absorbed in the chemical potential. Since the mass term breaks the sublattice symmetry, for $\Delta\neq 0$ we have $\Sigma_A \neq \Sigma_B$ and therefore $\Sigma_{A/B} = \frac{1}{2}\left[\Sigma_A + \Sigma_B \pm \left(\Sigma_A - \Sigma_B\right)\right]$.
The local self-energy can then be written in terms of
\begin{equation}
	\delta\Sigma \equiv \frac{\Sigma_B(\omega=0) - \Sigma_A(\omega=0)}{2} \geq 0
\end{equation}
as
\begin{equation}
	\Sigma_\mathrm{loc} = a\sigma_1 + b\sigma_2 - \delta\Sigma\;\sigma_3 = \begin{pmatrix}
	-\delta\Sigma & a - ib \\
	a + ib & \delta\Sigma
	\end{pmatrix}
	\label{eq:sigma_local}
\end{equation}
where $a,b\in \mathbb{R}$. With this we can express the complete self-energy (Eq.~\eqref{eq:sigma_loc_nonloc}
at $\omega = 0$) as
\begin{equation}
	\Sigma(k, \omega=0) = (a\sigma_1 + b\sigma_2) - \delta\Sigma\,\sigma_3 + \Sigma_\mathrm{non-loc}(k, \omega=0).
	\label{eq:sigma_three_parts}
\end{equation}
We have now made explicit the three terms leading to a shift of the phase transition in the topological Hamiltonian. We can readily see that the effect of the diagonal part (Hartree + corrections) of the {\it local} self-energy is proportional to $\sigma_3$ and therefore constitutes a mere shift of the mass term
\begin{equation}
	\Delta \mapsto \Delta - \delta\Sigma,
	\label{eq:renormalized_delta}
\end{equation}
which already describes the results obtained in many studies with both local
and non-local methods \cite{Vanhala2016,Tupitsyn2019,Budich2013}, as only the value of $\delta\Sigma$ varies slightly,
without changing the qualitative behavior. The negative shift of
Eq.~\eqref{eq:renormalized_delta} in $\Delta$ corresponds to a positive shift
of the topological phase transition along the $\Delta$ axis (see \figref{fig:phase_diagram}). 
Note that since in the local self-energy away from half filling (per
site) non-local corrections are present, the exact value of $\delta\Sigma$ is not reproduced by local diagrams only, e.g.~in DMFT.

We concentrate now on the off-diagonal contribution to $\Sigma_\mathrm{loc}$ (Eq.~\eqref{eq:sigma_local}).
For the terms proportional to $\sigma_1$ and $\sigma_2$, we cannot
simply write down a mapping like Eq.~\eqref{eq:renormalized_delta}, since such
a (constant) term does not appear as an individual parameter in the original
Hamiltonian. The only similar term in Eq.~\eqref{eq:haldane_hamiltonian_matrix}
is $t_1\sigma_1$, which originates from the coupling of $A$ and $B$ sites
within the unit cell. By using the general approach given by
Eq.~\eqref{eq:sigma_three_parts}, we can tune the model between coupled
one-dimensional chains ($a=-t_1$, $b=0$) and coupled dimers ($|a|\gg t_1$).
Interestingly, tuning the hopping beyond the chain model, $a < -t_1$, a novel
non-trivial phase with $C=-1$ appears at $\Delta<\Delta_c$, similarly to an
effect observed in the dimerized Hofstader model \cite{Lau2015}.  We note, however, that
if we restrict ourselves to the calculation of the self-energy for the
Haldane-Hubbard model through, for instance, the TPSC approach, the sign of the
local self-energy off-diagonal term is always positive.

\begin{figure}
	\includegraphics[scale=1]{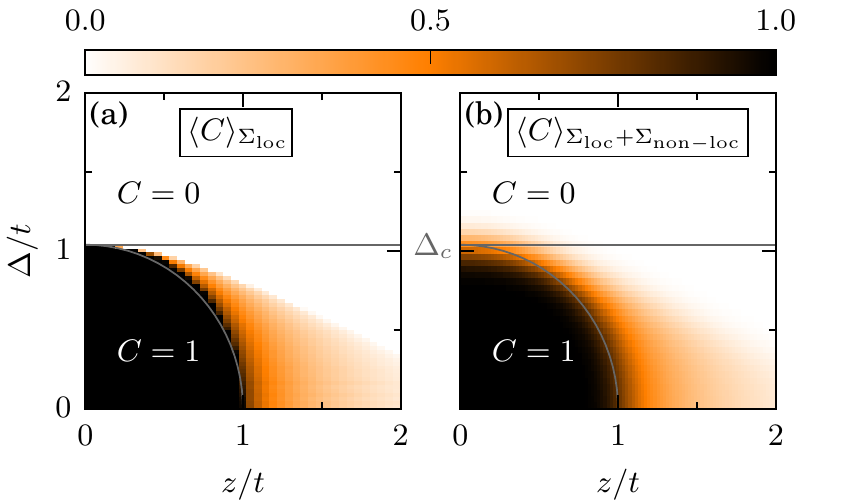}
	\caption{Average Chern number $\langle C\rangle$ as a function of $\Delta$ and $z$, Eq.~\eqref{eq:z_def}. (a) Only local terms are considered, (b) sampling procedure includes non-local contributions. We added lines marking the non-interacting phase transition (horizontal) and the shifted transition as a function of $z$. The non-trivial phase is stable in the black region and the phase transition with a local self-energy lies in the shaded region, i.e., the phase transition is shifted towards smaller $\Delta$ w.r.t.~the non-interacting case. This shift becomes significant at large $z/t$.}
	\label{fig:avg_chern_offdiagonal}
\end{figure}
We proceed by numerically studying the effects of such an off-diagonal term   by
computing the Chern number, Eq.~\eqref{eq:chern_invariant}, with the algorithm given in \cite{Fukui2005} as a function of
\begin{equation}
	z=|\Sigma_\mathrm{loc}^{AB}| = |a+ib|
	\label{eq:z_def}
\end{equation}
for a single spin. Since $H_t$ is diagonal in the spin basis,
keeping both spins would unnecessarily double the dimensionality of the
problem. Note that this implies a Chern number $C=1$ below the critical mass $\Delta_c$ instead of $C=2$ as in the spinful model.
In \figref{fig:avg_chern_offdiagonal}(a) we plot the average Chern number $\langle C\rangle$, where $\langle \ldots \rangle$ is the average over a number of samples with random complex phases of $\Sigma_\mathrm{loc}^{AB}$.
We find that upon perturbing the Hamiltonian with a constant off-diagonal term, the topological insulator is robust within a well-defined region (black). Depending on the value of the complex phase, the phase transition lies in the shaded region, which is located below the non-interacting phase boundary marked by $\Delta_c$, i.e.~the non-trivial phase region ($C=1$) generally shrinks.
This is a consequence of off-diagonal and diagonal parts of the
local self-energy having opposite effects on the topological phase (i.e.~down-/up-shift of the transition along the $\Delta$-axis), albeit the
diagonal contribution will typically be much larger for significantly large
$\Delta$. \figref{fig:avg_chern_offdiagonal}(b) shows, for comparison, the average Chern number
obtained by including in the calculation the non-local contributions to the self-energy and
will be discussed in  section \ref{sec:non-local}.

\subsection{Magnetic self-energy}
Before proceeding with the momentum-dependent (non-local) self-energy contributions, it is worthwhile to
analyze the effect of magnetism on the Chern number.
An odd total Chern number can only arise if the SU$(2)$ symmetry is spontaneously broken, i.e., in a magnetically ordered phase. This follows directly from the topological Hamiltonian, since the spins are decoupled and the Haldane-Hubbard Hamiltonian conserves SU$(2)$ symmetry. 
The mean-field equations are easily adapted to include an additional on-site magnetization $m$
\begin{equation}
	\Sigma_{\sigma}^\mathrm{MF} = \frac{U}{2}\left( n + \sigma m\right),
	\label{eq:sigma_mf}
\end{equation}
where $\sigma\in\{+1,-1\}$. Eq.~\eqref{eq:sigma_mf} is then rewritten in terms of $\delta n = (n_B-n_A)/2$ and $\delta m = (m_B-m_A)/2$ as
\begin{equation}
	\Sigma_{\sigma}^\mathrm{MF} = -\frac{U}{2} \left( \delta n \,\sigma_3 + \sigma \delta m \,\sigma_3 \right) + \mathrm{const}.
\end{equation}
In this description one identifies an additional spin-dependent renormalization of the mass term proportional to the magnetization difference $\delta m$. As in the SU$(2)$-symmetric case, an analogous calculation can be performed with the general self-energy.
In this case the mapping of Eq.~\eqref{eq:renormalized_delta} is modified to
\begin{equation}
	\Delta \mapsto \Delta - (\delta \Sigma + \delta \Sigma_\sigma),
\end{equation}
where the additional term is $\delta \Sigma_\sigma = \sigma (\Sigma_{B\uparrow} - \Sigma_{B\downarrow} - \Sigma_{A\uparrow} + \Sigma_{A\downarrow})/4$. Therefore, the two spins obtain different renormalizations, which can lead to one spin in the non-trivial phase ($C_\uparrow=1$) and the other in the trivial phase ($C_\downarrow=0$). The critical value is given by the condition
\begin{equation}
	\Delta - \Delta_c \leq \delta\Sigma + \delta\Sigma_{\sigma=+1},
\end{equation}
where $\Delta_c/t\approx 1.04$ marks the position of the non-interacting phase transition.

\section{Chern number analysis: Non-local contributions}
\label{sec:non-local}

In order to study the effect of the explicit momentum dependence of the self-energy on the Chern number
(the last term in Eq.~\eqref{eq:sigma_three_parts}), an analytic formula or parameterization of the self-energy would be helpful.
One such parameterization is possible within the Two-Particle Self-Consistent
method (TPSC) \cite{Vilk1994}, where the self-energy is parameterized by two
variables $U,U^\prime$, which are determined self-consistently. Within this
TPSC parametrization we did not detect any change of the Chern number in the Haldane-Hubbard model
with respect to the momentum-averaged TPSC.
Generalizing  the TPSC formula to an ansatz function that serves as a parameterized form of physical Green's functions
\begin{equation}
	\Sigma(k) = \Big(\big[V\!\left[U\right]+V\!\left[U^\prime\right]\big]\ast G^0\Big)(k),
	\label{eq:sigma_tpsc}
\end{equation}
where
\begin{equation}
	V[U] = \left(1-\chi^0U\right)^{-1}\chi^0,
\end{equation}
$\chi^0$ is the susceptibility and $U, U^\prime$ are the free parameters (here, $U, U^\prime$ depend on the site index $A, B$, i.e., there are four free parameters),
we do not find any topological phase transition induced by the momentum dependence of the self-energy while restricting ourselves to moderate values for $U,U^\prime$.

\subsection{Statistical study: Formalism}
Since we would like to systematically determine the importance of the momentum-dependence of the self-energy 
for a general class of interacting systems described by effective topological Hamiltonians,
we compute the statistical distribution of the Chern number across the space of possible self-energy functions beyond Eq.~\eqref{eq:sigma_tpsc}.

\begin{figure}
	\includegraphics[scale=1]{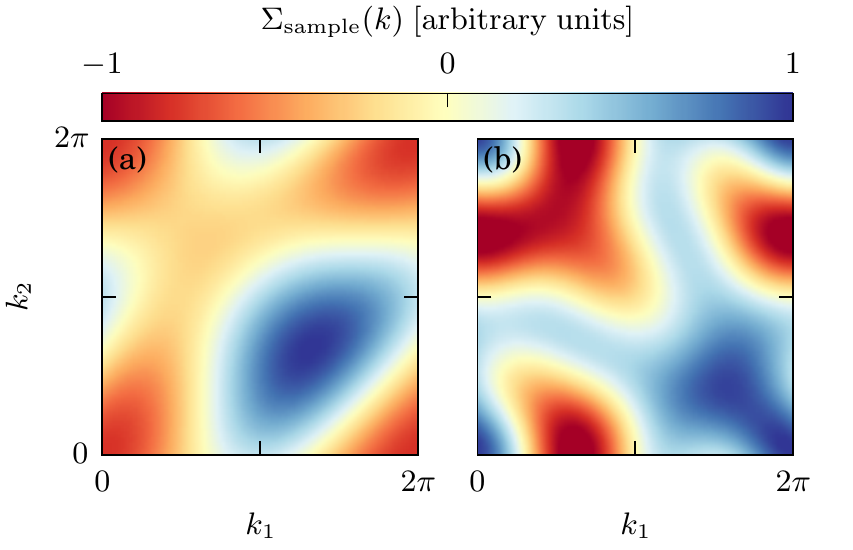}
	\caption{We compare the momentum-dependence of (a) a generic TPSC self-energy (here a fit where coefficients beyond $N_c=1$ vanish is shown) with (b) one of our random samples at $N_c=2$. Both functions follow the same symmetry constraints and are reasonably smooth.}
	\label{fig:sample_comp}
\end{figure}

The self-energy for the Haldane-Hubbard model is a complex $2\times 2$ matrix (for each spin)
and is block-diagonal in the spin space due to the absence of spin-mixing terms. Therefore, we focus on a single spin for this investigation as the task is easily separable and both spins are treated in exactly the same way. We define the following parameterization of the momentum-dependent part of the self-energy
\begin{equation}
	\Sigma_\mathrm{non-loc} = \begin{pmatrix}
		f_1 & f_2+if_3 \\
		f_2 - if_3 & -f_1
	\end{pmatrix},
	\label{eq:sigma_matrix_composition}
\end{equation}
which contains three independent real-valued periodic functions $f_1, f_2, f_3: \mathbb{R}^2\rightarrow \mathbb{R}$ and is hermitian by construction. The complete self-energy at $\omega=0$ is obtained from Eq.~\eqref{eq:sigma_three_parts}. The symmetry between the $\Sigma_{11}=f_1$ and $\Sigma_{22}=-f_1$ matrix elements is chosen in accordance with Eq.~\eqref{eq:sigma_diagonal_symmetry}. A generalization with $\Sigma_{22}=f_4$ would, however, be straightforward.
Further, we expand all functions $f_j$ in terms of Fourier components
\begin{equation}
\begin{split}
	f_j(k) = &\sum_{l_1, l_2, s} c_{s, l_1, l_2}\cos(l_1 k_1 + s l_2 k_2) \\
	&+ \sum_{l_1, l_2, s} c^\prime_{s, l_1, l_2}\sin(l_1 k_1 + s l_2 k_2),
\end{split}
	\label{eq:mc_function}
\end{equation}
where $j= 1, 2, 3$ and $s\in\{-1, 1\}, l_1,l_2\in\{0, ..., N_c\}$, $N_c$ being the order of the expansion. This expansion is convenient
due to the periodicity of the self-energy in momentum space. By sampling the
real expansion coefficients $c, c^\prime$ from a suitable probability
distribution we obtain samples of smooth self-energy functions. Due to the
completeness of the basis functions ($\sin,\cos$) the entire relevant space is
covered in the limit $N_c\rightarrow \infty$. We have verified that in order to
represent the TPSC or FLEX \cite{Bickers1989} self-energies with high accuracy one only needs $N_c=1$, see \figref{fig:sample_comp}(a).
At this low cutoff there are already sufficient degrees of freedom in
Eq.~\eqref{eq:mc_function} to sample a large variety of sensible functions. In
order to be more general, we increased the cutoff to $N_c=2$ and verified that
our sample functions do not oscillate unphysically, see \figref{fig:sample_comp}(b). In our calculations the qualitative results are independent of the choice of the cutoff, while an increase in the degrees of freedom generally leads to a decrease in the relative number of interesting samples (for which the Chern number is susceptible to $\Sigma_\mathrm{non-loc}$).
For the obtention of physical self-energies we chose
to sample the $c,c^\prime$ from a Normal distribution with zero mean and a
decaying standard deviation $\bar{\sigma}$
\begin{equation}
	\rho(c_{s,l_1,l_2}^{(\prime)}) = \mathrm{Normal}(\mu=0, \bar{\sigma}=\exp(-l_1-l_2)).
	\label{eq:prob_distribution}
\end{equation}
Due to the exponential decay of $\bar{\sigma}$ with the wavelength of the oscillation, the self-energies (Eq.~\eqref{eq:sigma_matrix_composition}) are guaranteed to be rather smooth. We verified that for instance a uniform distribution yields highly unsatisfactory samples, especially at larger cutoff.

Since the function samples of Eq.~\eqref{eq:mc_function} generally do not obey any spatial symmetries, we enforce certain symmetries of the Hamiltonian by adapting the sampling procedure. The sublattice symmetry, cf.~Eq.~\eqref{eq:sigma_diagonal_symmetry}, is already 
incorporated in Eq.~\eqref{eq:sigma_matrix_composition}.  General lattice
symmetries can be implemented on a higher level. In particular, applying a
symmetry operation to $f_j$ yields a constraint on the coefficients, which can
then be used to enforce the symmetry on the self-energy. In practice this
amounts to setting certain coefficients to zero or an interdependence between
some coefficients. For the Haldane-Hubbard model the diagonal elements have a mirror symmetry $M$
along the $k_2=-k_1$ axis, i.e., \begin{equation}
M :\;
	\begin{pmatrix}
		k_1\\
		k_2
	\end{pmatrix}
	\mapsto
	\begin{pmatrix}
		-k_2\\
		-k_1
	\end{pmatrix}.
\end{equation}

We now compute the weighted average Chern number on the space of differentiable
functions $\Sigma(k, \omega=0)$ as a function of their dispersion amplitudes
$d_a$ (Eq.\eqref{eq:da}). The average is weighted in nature, since we implement importance
sampling on the subspace of physical functions due to our choice of the distribution function $\rho$ (Eq.~\eqref{eq:prob_distribution}). After obtaining a sample for
the momentum-dependence we rescale the function $\Sigma_\mathrm{non-loc}$ to
the initially chosen dispersion amplitude $d_a$. The momentum-average $\sum_{k\in
1.\mathrm{BZ}} \Sigma_\mathrm{non-local}(k, \omega=0)$ vanishes by construction. The resulting
Chern number average $\langle C\rangle$ doubles as a standard deviation, since here the Chern
number can only take two values $C\in\{0, 1\}$, which square to themselves
($\mathrm{Var}[X] = \mathrm{E}[X^2] - \mathrm{E}[X]^2 = \mathrm{E}[X] -
\mathrm{E}[X]^2$). Therefore, the average Chern number can be interpreted as a
stochastic error measure. Since the expectation value does not accurately
describe the difference between the interacting and non-interacting system, it
is still not a sufficient measure for our statistical analysis. 

We have already found in Section \ref{sec:local} that a value $z>1$ can push the system into a $C=-1$ phase, which invalidates our earlier assumption that the Chern number is binary. Therefore, the average Chern number is an insufficient descriptor of the statistical distribution. A description in terms of probabilities of change is more appropriate. We define the probability for the Chern number to change w.r.t.~a reference Chern number $C_\mathrm{ref}$ as
\begin{equation}
	P(C\neq C_\mathrm{ref}) = \langle \min\{1, |C-C_\mathrm{ref}|\} \rangle,
\end{equation}
where $\langle \ldots\rangle$ is the sample mean. In fact, this definition is formally equivalent to the normalized distance between two probability distributions $d(X,Y)=\mathrm{E}_{X,Y}[|X-Y|]$ and therefore respects changes on a per-sample basis, which the average Chern number neglects. It is straightforward to show that this definition provides an appropriate measure for the probability of change in the sense that $P=0$ implies $C=C_\mathrm{ref}$ for all samples and $P=1 \Rightarrow C\neq C_\mathrm{ref}$. Additionally, $P$ is bounded to the interval $[0,1]$.

\subsection{Statistical study: non-local self-energy}
We now define $C_\mathrm{ref}=C_0$, which is the Chern number of the corresponding non-interacting system, and compute the probability $P(C\neq C_0)$.
\begin{figure}
	\includegraphics[scale=1]{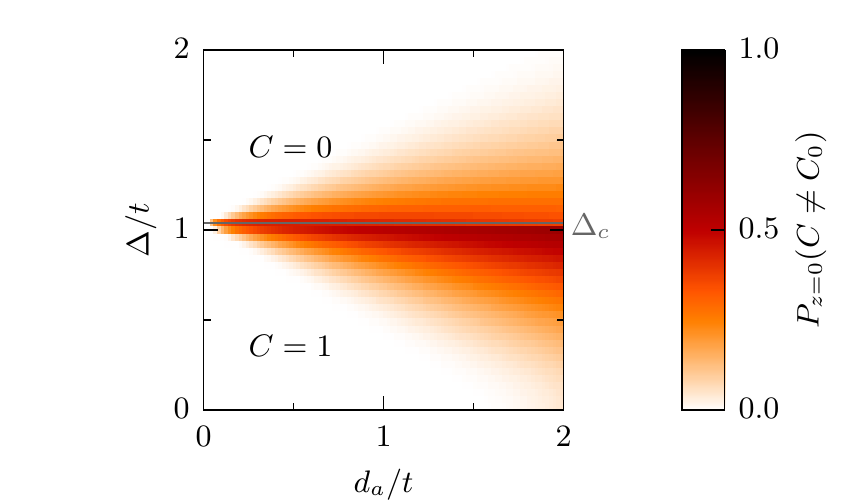}
	\caption{Probability $P_{z=0}(C\neq C_0)$ that the Chern number changes due to the full momentum-dependent self-energy compared to the non-interacting case for various values of $\Delta$ and $d_a$ at $z/t=0$. The position of the non-interacting phase transition $\Delta_c$ is marked by a gray line.}
	\label{fig:nonloc_diagram}
\end{figure}

In the simplified case, where we neglect the off-diagonal contributions to the local self-energy ($z=0$, Eq.~\eqref{eq:z_def}), we obtain a sharply peaked function shown in \figref{fig:nonloc_diagram},
which is centered around the phase transition at low $d_a$ and becomes
increasingly asymmetric for increasing $d_a$.  The smallness of
the stochastic error of the Chern number at small $d_a$ is due to
the peaked structure of the probability of change, which illustrates the
stability of the Chern number with respect to perturbations. At moderate to
large $d_a$, however, it turns out that the topologically non-trivial phase
$C=1$, which exists below $\Delta_c\approx 1.04$ at $U=0$, is more susceptible
to the addition of momentum dependent self-energies than the trivial insulator ($C=0$)
above $\Delta_c$. This means that on average, the effect of the momentum
dependence of the self-energy has the opposite sign as that of the local part, since it shifts the
transition towards lower $\Delta$ instead of larger $\Delta$.
Due to the distribution of finite probabilities around the local transition, we can regard the non-local contribution as a perturbation that leads to an uncertainty given by the spread of the probability distribution along the $\Delta$-axis. While this is rather small initially, a strong momentum-dependence of the self-energy can lead to a large uncertainty in the Chern number.

We note that the probabilities shown here depend on the sampling procedure used in our algorithm. From the rather physical nature of the restrictions to our sample functions, it is expected that in an unbiased average the probability of changing the Chern number would be rather low in all cases, since there is a large pool of functions which do not change the topology at all. We have probed the effect of restricting our trial function space of self-energies to only those functions satisfying the lattice symmetries of the non-interacting Hamiltonian and observed a small but noticeable increase in probabilities with enforced symmetries, which indicates that in a more general approach the unphysical samples lead to a decreased probability and therefore reduced contrast.

\subsection{Statistical study: total self-energy}
In the discussion so far we have neglected the off-diagonal terms of the local self-energy, which we have shown in \figref{fig:avg_chern_offdiagonal}(a) to have a comparatively weak impact on the Chern number, provided that $z$ is rather small. Now we add these terms back in by sampling the parameters $a,b$, cf.~Eq.~\eqref{eq:sigma_three_parts}. For this purpose we use the Euler representation of the off-diagonal value $a+ib = ze^{i\alpha}$ and sample the phase $\alpha$ from a uniform distribution $\alpha\in [0, 2\pi)$. The result is then a function of the absolute value $z$, which we have observed to contain the most relevant information. We compute the sample average over the Chern number, cf.~\figref{fig:avg_chern_offdiagonal}(b), which is remarkably similar to the one without the momentum-dependent part of the self-energy shown in \figref{fig:avg_chern_offdiagonal}(a). In fact, by comparing the average Chern numbers with and without the explicit momentum dependence we see that the effect of the momentum dependence is an additional uncertainty around the local result, which becomes broader for larger $d_a$ and is consistent with the result obtained without the off-diagonal terms of the local self-energy.

For our statistical analysis we distinguish between the relative probabilities $P(C\neq C_0)$, where $C_0$ is the Chern number of the non-interacting model, and $P(C\neq C_\mathrm{loc})$, where $C_\mathrm{loc}$ is the local Chern number computed from the topological Hamiltonian (Eq.~\eqref{eq:topological_hamiltonian_def}) with $\Sigma(k,\omega=0) = \Sigma_\mathrm{loc}$, cf.~Eq.~\eqref{eq:sigma_local}. The first probability characterizes the change with respect to the non-interacting case, while the second considers only the effect of the momentum-dependence of the self-energy. Since $d_a$ and $z$, characterizing the strength of the momentum-dependence and local self-energy, respectively, are both merely parameters in our model, these are simply different viewpoints onto the same problem, where the frame of reference is chosen differently to focus attention on only one parameter.

\begin{figure}
	\includegraphics[scale=1]{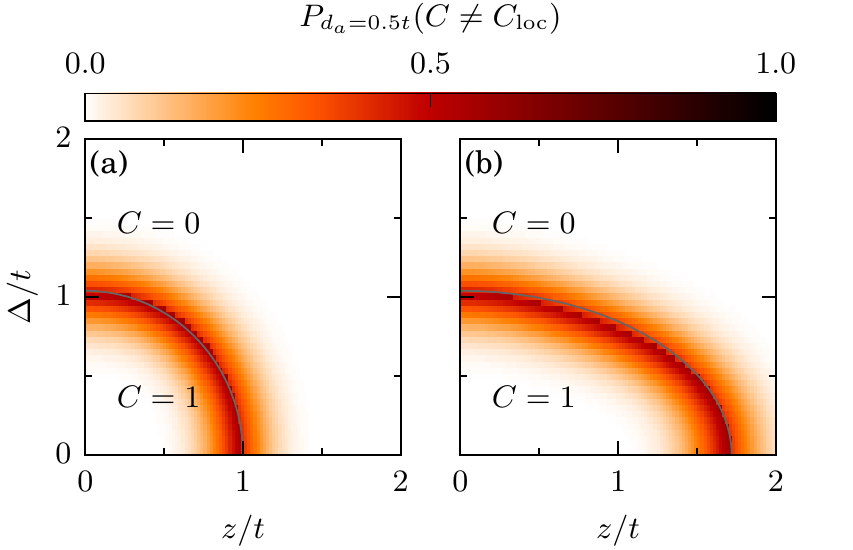}
	\caption{Probability $P_{d_a=0.5t}(C\neq C_\mathrm{loc})$ that the Chern number changes as an effect of the explicit momentum dependence $\Sigma_\mathrm{non-loc}(k)$ compared to the purely local case for two different phases of the local self-energy (a) $\alpha =0$ and (b) $\alpha =\pi/2$. Finite probabilities exist only around the local transition and decay with increasing the distance to the transition. The width of the finite-probability band depends on the dispersion amplitude $d_a$. Here, $d_a/t=0.5$.}
	\label{fig:statistical_uncertainty}
\end{figure}

We now focus our attention on the effect of the momentum-dependence. Since $\Sigma_\mathrm{loc}^{AB}=ze^{-i\alpha}$, in an average over different values of $\alpha$ one will automatically observe the changes due to a phase difference, which are unrelated to the momentum dependence. Hence, $P(C\neq C_\mathrm{loc})$ can only be computed as a function of the phase $\alpha$.
We have computed the probability $P(C\neq C_\mathrm{loc})$ that the momentum-dependent part of the self-energy $\Sigma_\mathrm{non-loc}$ changes the Chern number for many different values of $\alpha$ and show in \figref{fig:statistical_uncertainty} at two specific values $\alpha=0,\pi/2$ that a finite probability indeed exists only close to the respective local phase transition. We have verified that this is true independently of the phase $\alpha$ of the local self-energy. The probability can be described as a bell curve placed on top of each point on the local transition line. The width of this curve is proportional to the self-energy dispersion amplitude $d_a$ and coincides roughly with the result of \figref{fig:nonloc_diagram}.

As a result, the Chern number can actually be regarded as separable in the sense that the effects of the local and non-local terms in our representation of Eq.~\eqref{eq:sigma_loc_nonloc} are cumulative. This means that
\begin{equation}
	C = C_\mathrm{loc} + \delta C_\mathrm{non-loc},
	\label{eq:chern_number_linear}
\end{equation}
where the latter part is a random variable, whose probabilities for non-zero values decay with increasing distance to the local phase transition on a length scale proportional to $d_a$. The correction $\delta C_\mathrm{non-loc}$ is therefore---at low to intermediate $d_a$---only relevant relatively close to the local phase transition. Based on this observation, we can conclude that the topological phase diagram is well-described qualitatively by the local self-energy, while the explicit momentum-dependence only leads to a statistical error bar, the width of which can be inferred from \figref{fig:nonloc_diagram}.

\begin{figure}
	\includegraphics[scale=1]{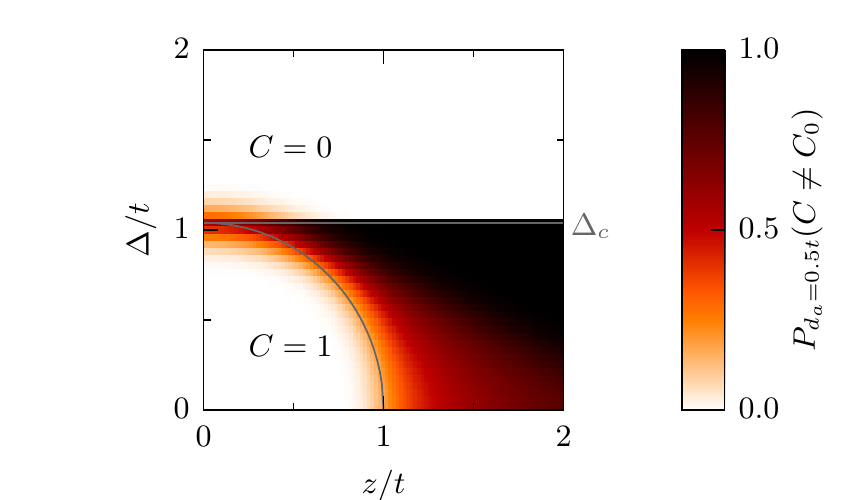}
	\caption{Probability $P_{d_a=0.5t}(C\neq C_0)$ that the Chern number changes due to the full momentum-dependent self-energy compared to the non-interacting case for various values of $\Delta$ and $z$ at $d_a/t=0.5$.}
	\label{fig:nonloc_diagram2}
\end{figure}

Finally, in \figref{fig:nonloc_diagram2} we show at a fixed value $d_a=0.5t$ the probability of change $P(C\neq C_0)$ with respect to the non-interacting case while considering the full self-energy, which largely resembles the result obtained for only the local self-energy, with an added uncertainty around the phase transition.

\section{Discussion}
In the following we want to emphasize the most important implications of our results.
\subsection{General Implications}
The Chern number as defined in Eq.~\eqref{eq:chern_invariant} is a direct measure of the momentum-dependence of the Hamiltonian. It is therefore expected that introducing a perturbation in the shape of an arbitrary function of momentum---in this case the self-energy---will have a large impact.

Our study reveals a paradox, where, in fact, the local perturbations have a much more immediate effect on the location of the topological phase transition, while the non-local contribution merely adds a rather small uncertainty around the local result. Therefore, the Chern number is really rather robust against non-local perturbations to the Hamiltonian.

\subsection{Discussion of the Phase Diagram}
Regarding the phase diagram of the Haldane-Hubbard model, cf.~\figref{fig:phase_diagram}, we draw the following conclusion. The $C=2$ to $C=0$ transition ($C=1$ to $C=0$ for each spin) is described very well by the local part of the self-energy, which is also reflected in the remarkable agreement between the DMFT and BDMC results. In fact, we have shown in an earlier publication \cite{Mertz2018} that in the presence of the mass $\Delta$ the momentum-dependence of the self-energy is rather weak for a wide range of parameters. Coincidentally, the TI$\rightarrow$BI transition lies within the non-dispersive regime, hence DMFT is expected to be very accurate.

The symmetry-broken phase with $C=1$, however, lies close to the Mott insulator, where the momentum-dependence plays a larger role. However, we note that we expect the momentum-dependence of $\Sigma$ to be the smaller contribution to the shift of the phase transition in comparison with BDMC, since it is closely related to the onset of a finite magnetization, which is predominantly reflected in the local self-energy.
Including  non-local (diagrammatic) corrections to the local self-energy should therefore produce a qualitatively correct phase diagram, while the non-local self-energy only leads to a small correction.

Our results are in principle applicable to other topological models, many of which contain a similar mass term, where, based on the published phase diagrams, we expect qualitatively similar results.

\begin{acknowledgments}
Most calculations were performed on the Goethe HLR high-performance computer of the Goethe Universit\"at. The authors would like to thank the Hessian Competence Center for High Performance Computing -- funded by the Hessen State Ministry of Higher Education, Research and the Arts. K.~Z.~and R.~V.~acknowledge financial support by the Deutsche Forschungsgemeinschaft through Grant No.~SFB/TR 49.
\end{acknowledgments}

\bibliographystyle{unsrt}
\bibliography{bibliography}

\end{document}